\documentclass[%
 reprint,
twocolumn,
superscriptaddress,
showpacs,preprintnumbers,showkeys,
 amsmath,amssymb,
 aps,
prc,
floatfix,
]{revtex4}

\usepackage{graphicx}% Include figure files
\usepackage{dcolumn}% Align table columns on decimal point
\usepackage{bm}% bold math
\usepackage{amssymb}
\usepackage{float}
\usepackage[caption = false]{subfig}
\usepackage{hyperref}% add hypertext capabiliti
\begin{document}

%\preprint{APS/123-QED}

\title{Background evaluations for the chiral magnetic effect  with normalized correlators using a multiphase transport model}
%\ifalse
%\affiliation[inst1]{Key Laboratory of Nuclear Physics and Ion-beam Application (MOE), and Institute of Modern Physics, Fudan University, Shanghai-200433, People’s Republic of China}
%\affiliation[inst2]{Department of Physics and Astronomy, University of California, Los Angeles, California 90095, USA}

\author{Subikash Choudhury}\email{subikash@fudan.edu.cn}
\affiliation{Key Laboratory of Nuclear Physics and Ion-beam Application (MOE), and Institute of Modern Physics, Fudan University, Shanghai-200433, People’s Republic of China}
\affiliation{Department of Physics and Astronomy, University of California, Los Angeles, California 90095, USA}

\author{Gang Wang}\email{gwang@physics.ucla.edu}
\affiliation{Department of Physics and Astronomy, University of California, Los Angeles, California 90095, USA}

\author{Wanbing He}
\affiliation{Key Laboratory of Nuclear Physics and Ion-beam Application (MOE), and Institute of Modern Physics, Fudan University, Shanghai-200433, People’s Republic of China}

\author{Yu Hu}
\affiliation{Key Laboratory of Nuclear Physics and Ion-beam Application (MOE), and Institute of Modern Physics, Fudan University, Shanghai-200433, People’s Republic of China}

\author{Huan Zhong Huang}
\affiliation{Key Laboratory of Nuclear Physics and Ion-beam Application (MOE), and Institute of Modern Physics, Fudan University, Shanghai-200433, People’s Republic of China}
\affiliation{Department of Physics and Astronomy, University of California, Los Angeles, California 90095, USA}

\date{\today}
%\fi
\iffalse
\author{Subikash Choudhury}
\email{subikash.choudhury@cern.ch}
\author{Debojit Sarkar}
\author{Subhasis Chattopadhyay}
\email{sub@vecc.gov.in}
\affiliation{Variable Energy Cyclotron Centre, HBNI, 1/AF Bidhan Nagar, Kolkata 700064, India}
%\affiliation{Bose Institute, Department of Physics and Center for Astroparticle Physics and Space Science (CAPSS),EN-80, Sector V, Kolkata-700091, India}
\date{\today}
\fi

%\corauth[cor]{Corresponding author.}
%\ead{sub.chattopadhyay@gmail.com}
%\address[label1]{Variable Energy Cyclotron Centre, 1/AF-Bidhannagar, Kolkata-700064, India}

\begin{abstract}
The chiral magnetic effect (CME) induces an electric charge  separation in a chiral medium along the magnetic field that is mostly produced by spectator protons in heavy-ion collisions.
The experimental searches for the CME, based on the charge-dependent angular correlations ($\gamma$), however, have remained inconclusive, because the non-CME background contributions are not well understood. Experimentally, the $\gamma$ correlators have been measured with respect to the second-order ($\Psi_{2}$) and the third-order ($\Psi_{3}$) symmetry planes, defined as
$\gamma_{112}$ and $\gamma_{123}$, respectively. The expectation was that with a proper normalization, $\gamma_{123}$ would provide a data-driven estimate for the background contributions in $\gamma_{112}$.
In this work, we calculate different harmonics of the $\gamma$ correlators using a charge-conserving
version of a multiphase transport (AMPT) model to examine the validity of the said assumption.
We find that the pure-background AMPT simulations do not yield
an equality in the normalized $\gamma_{112}$ and $\gamma_{123}$, quantified
by $\kappa_{112}$ and $\kappa_{123}$, respectively. 
Furthermore, we test another correlator, $\gamma_{132}$, within AMPT, and discuss the relation between different $\gamma$ correlators.
\end{abstract}

%\item{PACS numbers}
\pacs{}
%\begin{keyword}

\keywords{Chiral Magnetic Effect, Heavy-ion collisions} 

%\end{keyword}
%\end{frontmatter}

\maketitle

\section{Introduction}
Quantum chromodynamics (QCD), the underlying theory of strongly interacting quarks and gluons, is in general perceived as ${\cal P}$-even~\cite{intro_1}, where ${\cal P}$ denotes the parity symmetry . However, there are provisions within the theory that allow parity to be violated locally~\cite{intro_2, intro_3}.
The local parity violation in QCD entails vacuum fluctuations that create gluon fields
with non-zero topological charges~\cite{intro_4}. The interaction between these non-trivial topological gauge fields and mass-less chiral fermions (quarks) may create a domain with a local chirality imbalance, or unequal numbers of left- and right-handed quarks~\cite{intro_5}. Imprints of such a local chirality imbalance can be uncovered experimentally in the presence of a strong magnetic field ($B$), which generates an  electric charge separation along the $B$ direction, referred to as the chiral magnetic effect (CME) \cite{intro_6, intro_7}.

The high-energy heavy-ion collisions at Relativistic Heavy Ion Collider (RHIC) 
and the Large Hadron Collider (LHC) are known to create conditions  conducive to the experimental detection of the CME. A deconfined QCD medium produced in the collision allows for the creation of the metastable ${\cal P}$-odd domains, while the energetic spectator protons generate a very strong $B$ field, with the initial magnitude on the order of 10$^{14}$ Tesla. The CME coupling between the ${\cal P}$-odd domains and the strong $B$ field will lead to different preferential emissions for different charges along the $B$ field or across the reaction plane (RP) of the system \cite{ intro_8, intro_9}. The RP is spanned by the impact parameter and the beam momenta .

Over the last decade, the experimental search for the CME has become a major objective for the heavy-ion research program
at RHIC and the LHC. See Refs.~\cite{intro_9a, intro_9b} for recent reviews. A commonly used observable of the CME-induced charge separation is 
the charge-dependent two-particle azimuthal correlator relative to the RP, defined as~\cite{intro_10}
\begin{equation}
 \gamma_{112} = \langle \cos( \phi_{\alpha} + \phi_{\beta} - 2\Psi_{\rm RP} ) \rangle ,
\end{equation}
where $\phi_{\alpha}$ and $\phi_{\beta}$ are the azimuthal angles of two particles, bearing same or opposite electric charges,
and $\Psi_{\rm RP}$ is the reaction plane angle.
In practice, $\Psi_{\rm RP}$ can be approximated by an event plane ($\Psi_{\rm EP}$) or the second-order symmetry plane ($\Psi_{2}$), estimated from the azimuthal distribution of final-state particles. The correlator thus obtained will then be corrected with the corresponding event plane resolution ($Res\{\Psi_2\}$) \cite{intro_10a}. Alternatively, $\gamma_{112}$ can be calculated with a reference particle $\phi_{c}$ instead of $\Psi_{2}$: 
\begin{eqnarray}
 \gamma_{112} &=& \langle \cos( \phi_{\alpha} + \phi_{\beta} - 2\Psi_{2} ) \rangle / Res\{\Psi_2\} \\
 &=& \langle \cos( \phi_{\alpha} + \phi_{\beta} - 2\phi_{\rm c} ) \rangle / v_{2,c},
\end{eqnarray}
where $v_{n,c}$ is the $n^{\rm th}$-order anisotropic flow of particle $c$,
\begin{equation}
v_n = \langle \cos[n(\phi - \Psi_n)]\rangle / Res\{\Psi_n\}.
\end{equation}
Conventionally, $v_2$ is called elliptic flow, and $v_3$, triangular flow.

The CME-induced charge separation will give rise to
a positively finite value of
$\Delta\gamma_{112}$ ($\equiv\gamma^{\rm OS}_{112} - \gamma^{\rm SS}_{112} $), where $\rm SS$ means same-sign ($\alpha, \beta = +,+  ~\rm{or}~ -,-$), and $\rm OS$ denotes opposite-sign ($\alpha, \beta = +,- ~\rm{or}~ -,+$).
The measurements of $\Delta\gamma_{112}$ at RHIC and the LHC indeed have presented evidence that resembles the typical CME expectations \cite{intro_11, intro_12, intro_13, intro_13b, intro_14, intro_15, intro_16}.
However, the observed trend in data are also qualitatively compatible with non-CME background contributions, such as transverse momentum conservation (TMC), local charge conservation (LCC) and resonance decays, coupled with anisotropic flow and/or few-body non-flow correlations~\cite{intro_17, intro_18, intro_19, intro_20}. This has essentially prevented
the interpretation of the charge separation in data as a clear manifestation of the CME. Detailed evaluations of background contributions are warranted to advance experimental searches for the CME in heavy-ion collisions. 

Recently, the CMS collaboration proposed to estimate quantitatively the background in the charge separation observable $\gamma_{112}$ by measuring charge-dependent azimuthal correlations with respect to the third-order symmetry plane, $\Psi_{3}$. Since the pertinent CME signals are generated across the RP or $\Psi_{2}$, the charge separation relative to $\Psi_{3}$ only contains background contributions \cite{intro_15a,intro_16}. Similar to Eqs. (2) and (3), one defines the azimuthal correlator relative to $\Psi_{3}$,
\begin{eqnarray}
 \gamma_{123} &=& \langle \cos( \phi_{\alpha} + 2\phi_{\beta} - 3\Psi_{3} ) \rangle / Res\{\Psi_3\} \\
 &=& \langle \cos( \phi_{\alpha} + 2\phi_{\beta} - 3\phi_{\rm c} ) \rangle / v_{3,c}.
\end{eqnarray}
It was argued that the absence of the CME should validate an equality between the normalized charge-separation observables across $\Psi_{2}$
and $\Psi_{3}$:
\begin{equation}
\frac{\Delta \gamma_{112} } {v_2\Delta \delta } \approx \frac{\Delta \gamma_{123} } {v_3\Delta \delta}   ~\rm{or}~ \kappa_{112} \approx \kappa_{123}.
\end{equation}
Here, $\Delta \delta = \delta^{\rm OS} - \delta^{\rm SS}$, and $\delta = \langle \cos( \phi_{\alpha} - \phi_{\beta} ) \rangle$.

Such an equality was roughly supported by the CMS measurements, seemingly presenting a challenge to the CME interpretation of the observed charge separation in Pb+Pb collisions at 5.02 TeV. 
Preliminary STAR data also demonstrate that $\kappa_{112}$ and $\kappa_{123}$ are both around 2 for most centrality bins in Au+Au collisions at 200 GeV \cite{intro_16, intro_16a}.
However, before using $\kappa_{123}$ as a data-driven background estimate for $\kappa_{112}$, one should examine the aforementioned equality with a realistic background-only model. 
In the present work, we perform an explicit test of this idea with a multiphase transport model (AMPT) \cite{ampt_1},
which only contains non-CME backgrounds. Furthermore, we also  investigate another variant of the $\gamma$ correlator, 
\begin{eqnarray}
 \gamma_{132} &=& \langle \cos( \phi_{\alpha} - 3\phi_{\beta} + 2\Psi_{2} ) \rangle / Res\{\Psi_2\} \\
 &=& \langle \cos( \phi_{\alpha} - 3\phi_{\beta} + 2\phi_{\rm c} ) \rangle / v_{2,c}.
\end{eqnarray}
$\gamma_{132}$ employs the same second-order event plane as $\gamma_{112}$, and provides a first-order background estimate for $\gamma_{112}$. Similar to $\kappa_{112}$ and $\kappa_{123}$, we define
\begin{equation}
\kappa_{132} = \frac{\Delta\gamma_{132}}{v_2\Delta\delta}.
\end{equation}
A more detailed discussion on the relation between these three $\gamma$ correlators can be found in Sec. IV and in Appendix A.

\section{The AMPT Model}
AMPT is a hybrid transport event generator that describes different stages of a heavy-ion collision at relativistic energies.
This model has four major steps: the initial conditions, the partonic evolution, the hadronization,
and the hadronic interactions. For the initial conditions, AMPT uses the
spatial and momentum distributions of minijet partons and excited soft strings, adopted in the Heavy Ion Jet Interaction Generator (HIJING) \cite{ampt_2}.
Then Zhang's parton cascade (ZPC) \cite{ampt_3}
is exploited to manage the partonic evolution, characterized
by two-body parton-parton elastic scattering with the parton interaction cross section
obtained from pQCD calculations: ${\sigma_{p}\simeq 9\pi\alpha_{s}^{2}/2\mu^{2}}$.
Here $\alpha_{s}$ is the QCD coupling constant for strong interactions, and 
$\mu$ is the Debye screening mass of gluons in the QGP medium. At the end of the partonic evolution,
the spatial quark coalescence is implemented to achieve quark-hadron phase transition in the string melting (SM) version of AMPT.
In this approach,
spatially close quark-antiquark pairs (triplets) are recombined to form mesons (baryons).
Finally, the hadronic interactions are modelled by A Relativistic Transport calculations (ART)~\cite{ampt_4}.

The SM version of AMPT
reasonably well reproduces particle spectra and elliptic
flow in Au+Au collisions at 200 GeV and Pb+Pb collisions at 2.76 TeV.
In this study, the SM v2.25t4cu of AMPT has been used to simulate Au+Au collisions at 200 GeV.
This version assures charge conservation, which is particularly important for the CME-related studies.
We set the parton scattering cross section to 3 mb, the strong coupling constant ($\alpha_{s}$) to 0.33, and the Debye screening mass ($\mu$) to 2.265 fm$^{-1}$. 
The parameters for the Lund string fragmentation function~\cite{ampt_5},
\begin{equation}
 f(z) \propto (1-z)^{a}\exp(-bm_{T}^{2}/z),
\end{equation}
are kept as $a=0.55$ and $b=0.15$ GeV$^{-2}$, where $z$ denotes the light cone momentum fraction.

\section{Analysis}
In our analyses of the AMPT events, the centrality intervals are defined by slicing the impact parameter distribution.
For all the observables, the particles of interest (POI) come from midrapidities ($|\eta| < 1$) with transverse momentum $0.2 < p_{T} <  2$ GeV/$c$.
The $n^{\rm th}$-order event plane (EP) is reconstructed with a broad pseudorapidity
spectrum of charged particles in the range of $|\eta| < 4.5$:
\begin{equation}
 \Psi_{n} = \frac{1}{n} \tan^{-1}[ \frac{\sum \omega_{i} \sin(n\phi_{i})}{\sum \omega_{i} \cos(n\phi_{i})} ],
\end{equation}
where $\phi_i$ is the azimuthal angle of particle $i$, and $\omega_i$ is its weight. The weight in units of GeV/$c$ is chosen to
be linear with $p_T$ up to 2 GeV/$c$.
Note that in the calculations of $v_n$ and $\gamma$,  self-correlation has been removed to prevent POI from contributing to $\Psi_{n}$.

Besides the full event plane method, we also explore the scalar product (SP) approaches with $\eta$ gaps between sub-events and POI to mitigate nonflow correlations. 
In this method, $v_{n}$ is calculated with
\begin{equation}
 v_{n}^{A(B)} \{\rm SP\} = \frac{ \langle Q_{n}^{\rm POI}Q_{n}^{*A(B)} \rangle}{\sqrt{\langle Q_{n}^{A}Q_{n}^{*B} \rangle} }.
\end{equation}
$Q_{n}^{A}$ and $Q_{n}^{B}$ are the $n^{\rm th}$-order complex flow vector,
evaluated from sub-events $A$ and $B$ in the $\eta$ ranges of $-5.1 < \eta <  -2.1$ and $2.1  < \eta <  5.1$, respectively.
Q$_{n}^{\rm POI}$ refers to the flow vector for POI with $|\eta| < 1$.
We take an average of $v_{n}^{A}$ and $v_{n}^{B}$ to be the final $v_{n}\{\rm SP\}$.

\section{Results}
Since the non-CME backgrounds
in the $\gamma$ correlators can only exist in the presence of anisotropic flow, we want to first check $v_2$ and $v_3$ from the AMPT simulations. 
Figure 1 compares
$v_{2}$ (upper panel) and $v_{3}$ (lower panel) between
AMPT calculations and STAR data \cite{STAR_v2_2005, STAR_v3_2013} in Au+Au collisions at $\sqrt{s_{NN} } = 200$ GeV.
With the event plane method, AMPT can describe STAR's $v_{2}$ and $v_{3}$ reasonably well.
Within AMPT, the $\eta$ gap introduced in $v_2\{\rm SP\}$ only causes small deviations from the $v_2\{\rm EP\}$ results.
In the case of $v_3$, 
the two approaches demonstrate
larger differences, which may arise from nonflow, flow fluctuation and longitudinal flow decorrelation.

\begin{figure}[htbp]
\centering
\includegraphics[scale=0.40,keepaspectratio]{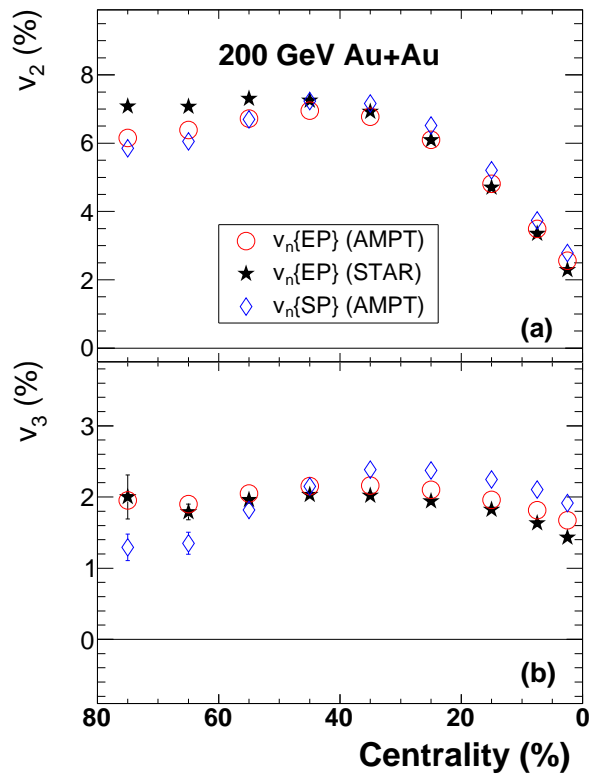}
\caption{(color online) Centrality dependence of elliptic flow, $v_{2}$  (upper) and triangular flow, $v_{3}$ (lower)
for AMPT and STAR results \cite{STAR_v2_2005, STAR_v3_2013} in Au+Au collisions at $\sqrt{s_{NN} } = 200$ GeV. Flow harmonics from AMPT are calculated with the event plane and the scalar product methods.}
\label{v2Cent}
\end{figure}

The CME-induced charge separation, as well as all the background sources, such as TMC, LCC and resonance decays, can be reflected in the two-particle correlations,
which makes $\delta$ an important physics observable.
In the following discussions and equations, we will approximately regard the $\delta$-related terms as pure backgrounds, because $\delta$ is dominated by background contributions, and $\delta$ is  always coupled with $v_2$ or $v_3$, which further suppresses the CME signal by over an order of magnitude. 
Figure 2 shows AMPT calculations of $\delta^{\rm OS(SS)}$ (upper) and $\Delta\delta$ (lower) as functions of centrality in Au+Au collisions at $\sqrt{s_{NN} } = 200$ GeV. The STAR results \cite{intro_12} are also plotted in comparison.
Although $\delta^{\rm OS}$ and $\delta^{\rm SS}$ from AMPT seem to display different trends from the corresponding STAR data, $\Delta\delta$ reveals a good consistency between AMPT simulations and STAR measurements.
As the major backgrounds in the $\gamma$ correlators result from the coupling between collective motion and two-particle correlations, the consistency shown in Figs. 1 and 2 makes AMPT a promising candidate for the background estimation for the $\gamma$ correlators.

Note that $\delta^{\rm OS}$ and $\delta^{\rm SS}$ shares a mutual background due to collective motion and momentum conservation \cite{intro_19}. This is also true for $\gamma^{\rm OS}$ and $\gamma^{\rm SS}$. For example, momentum conservation tends to push the two POI back to back, shifting both $\delta^{\rm OS}$ and $\delta^{\rm SS}$ downwards. The difference in this mutual background between AMPT and STAR data indicates a weaker manifestation of momentum conservation in real data, which may be subject to the  detector details.
The focus of this work, however, is the difference between opposite-sign and same-sign correlations, which is a more robust observable. 

\begin{figure}[tbp]
\centering

\includegraphics[scale=0.40,keepaspectratio]{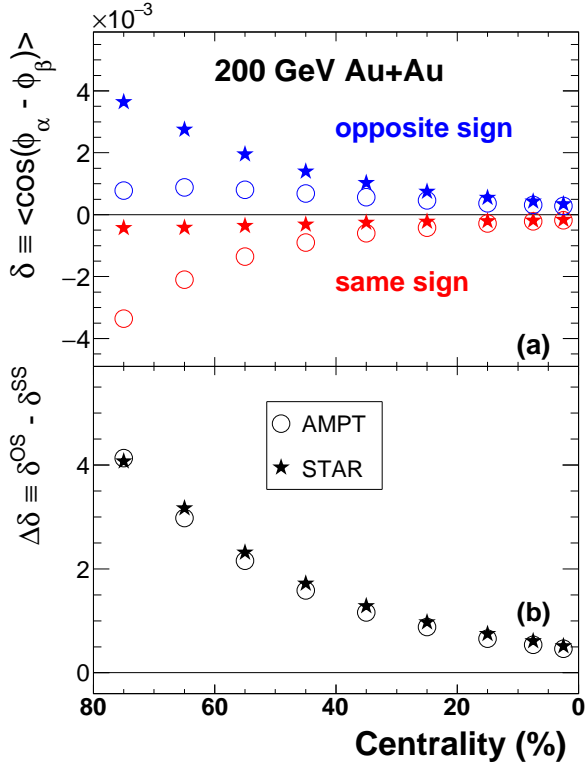}
\caption{(color online) Two-particle correlations $\delta^{\rm SS}$, $\delta^{\rm OS}$ (upper) and $\Delta\delta$ (lower) as functions of
centrality for AMPT and STAR results \cite{intro_12} in Au+Au collisions at $\sqrt{s_{NN} } = 200$ GeV.}
\label{v1cent}
\end{figure}

Figure 3 depicts the centrality dependence of $\gamma_{112}$ (upper), $\gamma_{132}$ (middle) and $\gamma_{123}$ (lower) for same-sign and opposite-sign
particle pairs from AMPT in Au+Au collisions at 200 GeV. The event plane method utilizes the charged particles with $|\eta|<4.5$ to reconstruct the event plane, and there is no $\eta$ gap between the two POI. As a systematic check, the 3-particle approach takes the reference particle from $2.1<|\eta|<5.1$, and introduces an extra $\eta$ gap of 0.4 between the two POI
to reduce short-range nonflow backgrounds. Note that this $\eta$ gap will modify the kinematic region of the POI. All the $\gamma$ correlators exhibit a clear charge dependence over the
centrality range under study. The two approaches result in very similar $\gamma_{132}^{\rm SS}$ or $\gamma_{132}^{\rm OS}$,
but it is not the case for $\gamma_{112}$ and $\gamma_{123}$. The observed difference can be attributed to the different kinematic regions of the POI,
which affects the behaviours of momentum conservation.

\begin{figure}[tbp]
\centering
\includegraphics[scale=0.45,keepaspectratio]{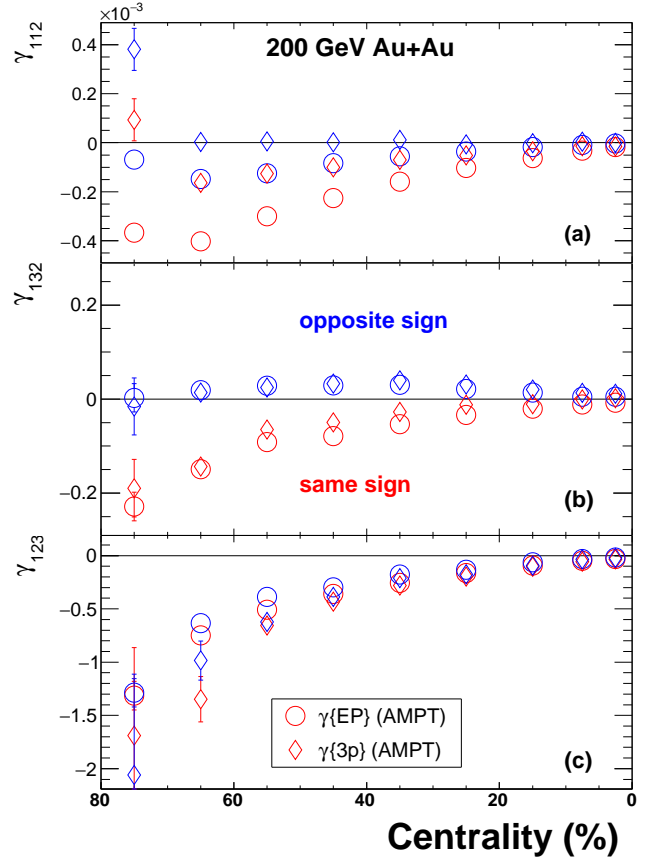}
\caption{(color online) Centrality dependence of $\gamma_{112}$ (upper), $\gamma_{132}$ (middle) and $\gamma_{123}$ (lower) for same-sign and opposite-sign
particle pairs, from AMPT in Au+Au collisions at $\sqrt{s_{NN} } = 200$ GeV. Both event plane and 3-particle approaches are displayed.}
\label{Gmnk_Cnt}
\end{figure}

In order to cancel out the mutual background, we present in Fig.~4 $\Delta\gamma_{112}$ (upper), $\Delta\gamma_{132}$ (middle) and $\Delta\gamma_{123}$ (lower) from AMPT in Au+Au collisions at 200 GeV. The two approaches show very similar results in all the three correlations, with the 3-particle magnitudes slightly lower. Remarkably, in this complete background scenario of AMPT,
a sizeable charge separation can be observed across all the harmonics of the $\gamma$ correlators. Even the centrality dependence of these $\Delta\gamma$ correlations qualitatively resembles
the one expected by the CME picture. In comparison, the STAR data of $\Delta\gamma_{112}$ \cite{intro_12} are also shown, with  magnitudes significantly larger than those from AMPT.
The background contributions from AMPT alone cannot explain $\Delta\gamma_{112}$ measured by STAR.

\begin{figure}[htbp]
\centering
\includegraphics[scale=0.45,keepaspectratio]{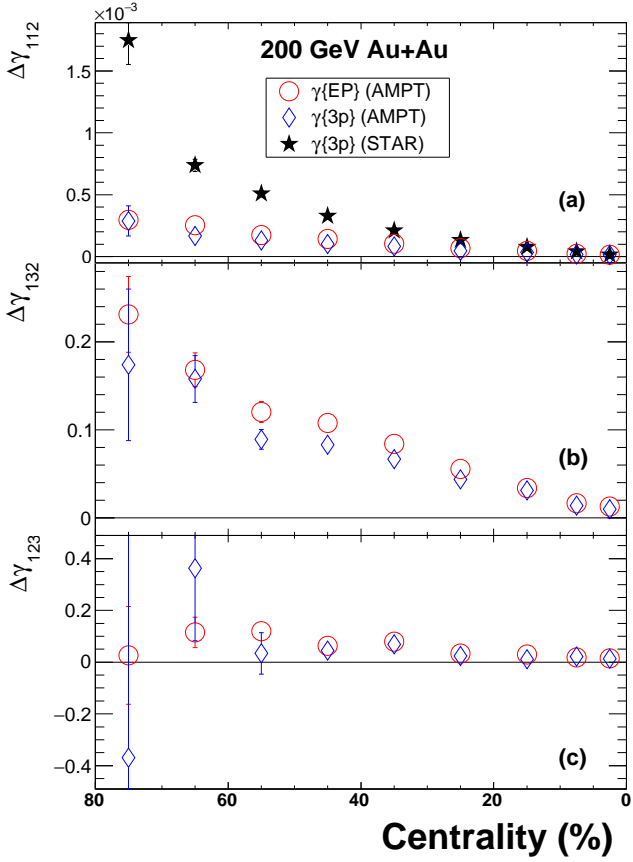}
\caption{(color online) Centrality dependence of $\Delta\gamma_{112}$ (upper), $\Delta\gamma_{132}$ (middle) and $\Delta\gamma_{123}$ (lower) from AMPT in Au+Au collisions at $\sqrt{s_{NN} } = 200$ GeV. Both event plane and 3-particle approaches are displayed. STAR data of $\Delta\gamma_{112}$ \cite{intro_12} are also shown in panel (a) for comparison.}
\label{DGmnk_Cnt}
\end{figure}

Normalized observables, such as $\kappa_{112}$, $\kappa_{132}$ and $\kappa_{123}$, can facilitate the comparisons between different collision systems,
different beam energies and different model implementations.  
Figure 5 delineates the centrality dependence of $\kappa_{112}$ (upper), $\kappa_{132}$ (middle) and $\kappa_{132}$ (lower) from AMPT in Au+Au collisions at $\sqrt{s_{NN} } = 200$ GeV.
Results from the event plane and the 3-particle approaches are consistent with each other, and the following discussions will focus on the event plane method. The first thing to note is that $\kappa_{132}$ is very close to unity.
A constant fit over $0-80\%$
centrality range gives $\kappa_{132} = 0.98 \pm 0.03$. We want to invoke the cumulant notation to understand this. The cumulant represents the ``true" correlation between two quantities ($a$ and $b$), and is denoted by the double bracket
\begin{equation}
\langle \langle a\cdot b \rangle \rangle \equiv \langle a\cdot b \rangle - \langle a  \rangle \cdot \langle b \rangle.    
\end{equation}
Then $\gamma_{132}$ can be expanded in the following way
\begin{eqnarray}
\gamma_{132} &=& \langle \cos( \phi_{\alpha} - 3\phi_{\beta} + 2\Psi_{\rm RP} ) \rangle \nonumber \\
&=& \langle \cos(\phi_{\beta} - \phi_{\alpha} + 2\phi_{\beta} - 2\Psi_{\rm RP} ) \rangle \nonumber \\
&=& \langle \cos(\phi_{\beta} - \phi_{\alpha})\cos(2\phi_{\beta} - 2\Psi_{\rm RP} )  \rangle \nonumber \\
& &- \langle\sin(\phi_{\beta} - \phi_{\alpha})\sin(2\phi_{\beta} - 2\Psi_{\rm RP} )  \rangle  \\
&=& \delta\cdot v_2 + \langle \langle \cos(\phi_{\beta} - \phi_{\alpha})\cos(2\phi_{\beta} - 2\Psi_{\rm RP} )  \rangle \rangle \nonumber \\
& &- \langle \langle \sin(\phi_{\beta} - \phi_{\alpha})\sin(2\phi_{\beta} - 2\Psi_{\rm RP} )  \rangle\rangle.
\end{eqnarray}
From Eq.(15) to Eq.(16), we utilize the cumulant, and  $\langle\sin(2\phi_{\beta} - 2\Psi_{\rm RP})\rangle$ is zero because of symmetry.
The observation that $\kappa_{132}\approx1$ implies that the two cumulant terms in Eq.(16) tend to cancel each other.
If that is the case, then $\gamma_{132}^{\rm SS}$ and $\gamma_{132}^{\rm OS}$ should be close to $v_2\cdot\delta^{\rm SS}$ and $v_2\cdot\delta^{\rm OS}$, respectively. This expectation is indeed supported by the AMPT simulations in the upper panel of Fig.~6.
Following the same speculation, we predict in the lower panel of Fig.~6 the centrality dependence of $\gamma_{132}^{\rm SS(OS)}$
for experimental data in Au+Au collisions at 200 GeV
using $v_2\cdot\delta^{\rm SS(OS)}$ from STAR.
The cancellation of the two cumulant terms also makes $\gamma_{132}$ a nearly-pure-background observable. 

\begin{figure}[htbp]
\centering
\includegraphics[scale=0.45,keepaspectratio]{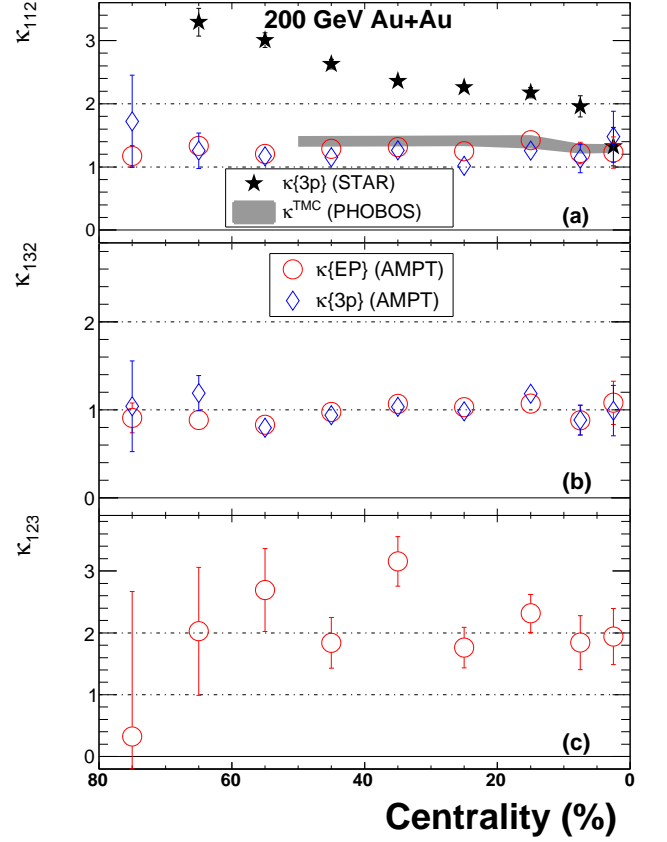}
\caption{(color online) Centrality dependence of $\kappa_{112}$ (upper), $\kappa_{132}$ (middle) and $\kappa_{123}$ (lower) from AMPT in Au+Au collisions at $\sqrt{s_{NN} } = 200$ GeV. The $\kappa_{123}$ results from the 3-particles method are not shown to avoid clutter. The STAR data of $\kappa_{112}$ and a background estimate using PHOBOS $v_2$ results are also shown in panel (a) in comparison.}
\label{v2pt}
\end{figure}

\begin{figure}[htbp]
\centering
\includegraphics[scale=0.45,keepaspectratio]{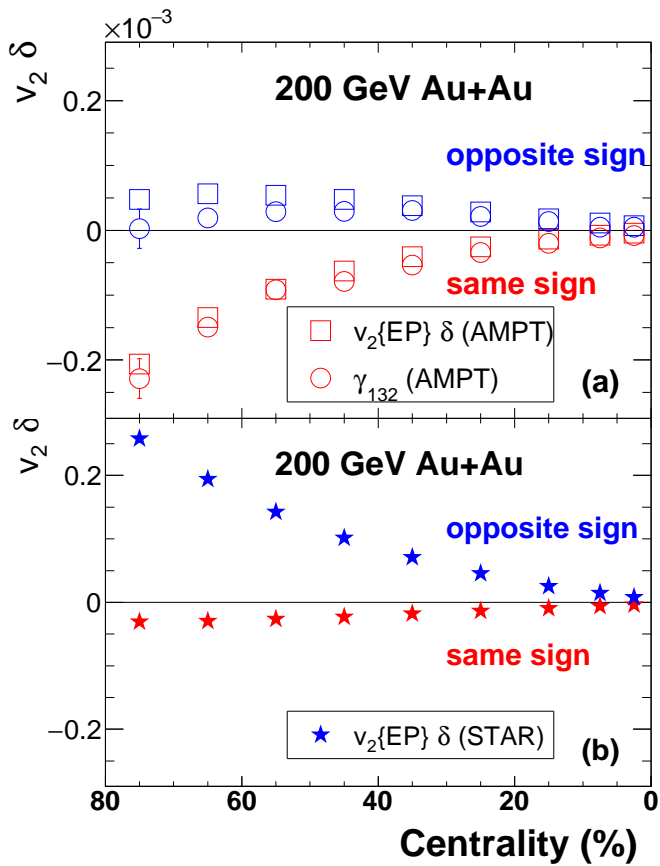}
\caption{(color online) Centrality dependence of $v_2\delta$ from AMPT (upper)  and from STAR data (lower) in Au+Au collisions at $\sqrt{s_{NN} } = 200$ GeV. The $\gamma_{132}$ results from AMPT are also shown in comparison.
The $v_2\delta$ values in the lower panel come from the STAR measurements of $v_2$~\cite{STAR_v2_2005} and $\delta$~\cite{intro_12}, and serve as predictions for $\gamma_{132}^{\rm OS(SS)}$ in data.}
\label{v2pt}
\end{figure}

An expansion of $\gamma_{112}$ similar to Eq.(16) reveals
\begin{eqnarray}
\gamma_{112} &=& \langle \cos( \phi_{\alpha} + \phi_{\beta} - 2\Psi_{\rm RP} ) \rangle \nonumber \\
&=& \delta\cdot v_2 + \langle \langle \cos(\phi_{\beta} - \phi_{\alpha})\cos(2\phi_{\beta} - 2\Psi_{\rm RP} )  \rangle \rangle \nonumber \\
& &+ \langle \langle \sin(\phi_{\beta} - \phi_{\alpha})\sin(2\phi_{\beta} - 2\Psi_{\rm RP} )  \rangle\rangle.
\end{eqnarray}
Now the two cumulant terms add up, instead of cancel out, making $\kappa_{112}$ deviate from unity, as shown in the upper panel of Fig.~5.
A constant fit over $0-80\%$
centrality range gives $\kappa_{112} = 1.29 \pm 0.03$ from the pure-background scenario in AMPT. The grey band in Fig.~5(a) displays an alternative estimate of the background due to collective motion and TMC.
%The background contribution in $\gamma_{112}$ was originally estimated for the ``flowing cluster" case~\cite{SVolo_3p}: $v_{2,{\rm cl}} \frac{\mean{\cos(\phi_\alpha+\phi_\beta-2\phi_{\rm cl})}}{\mean{\cos(\phi_\alpha-\phi_\beta})}$, where $\phi_{\rm cl}$ is the cluster emission azimuthal angle, and $\phi_\alpha$ and $\phi_\beta$ are the azimuthal angles of two decay daughters. The flowing cluster can be generalized to a larger portion of or even the whole event, through the mechanisms of TMC and/or LCC.
The TMC effect leads to the following pertinent correlation terms in $\delta$ and $\gamma_{112}$~\cite{Flow_CME}:
\begin{eqnarray}
\Delta\delta^{\rm TMC} &\propto& \frac{1}{N} 
\frac{{p_{T}}^{2}_{\rm \Omega}}{{p_T^2}_{\rm F}} 
\frac{1+({\bar v}_{2,{\rm \Omega}})^2-2{\bar{\bar v}}_{2,{\rm F}}{\bar v}_{2,{\rm \Omega}}} {1-({\bar{\bar v}}_{2,{\rm F}})^2},
\\
\Delta\gamma_{112}^{\rm TMC} &\propto& \frac{1}{N}   
\frac{{p_T}^2_{\rm \Omega}}{{p_T^2}_{\rm F}}
\frac{2{\bar v}_{2,{\rm \Omega}}-{\bar{\bar v}}_{2,{\rm F}}-{\bar{\bar v}}_{2,{\rm F}}({\bar v}_{2,{\rm \Omega}})^2} {1-({\bar{\bar v}}_{2,{\rm F}})^2}   
\nonumber \\
&\approx& \kappa_{112}^{\rm TMC} \cdot v_{2,{\rm \Omega}} \cdot \Delta\delta^{\rm TMC},
\end{eqnarray}
where $\kappa_{112}^{\rm TMC} = (2{\bar v}_{2,{\rm \Omega}}-{\bar{\bar v}}_{2,{\rm F}})/v_{2,{\rm \Omega}}$,
and ${\bar v}_{2}$ and ${\bar{\bar v}}_{2}$ represent the $p_T$- and $p_T^2$-weighted moments of $v_2$, respectively.
The subscript ``F" denotes an average of all produced particles in the full phase space; 
the actual measurements will be only in a fraction of the full space, denoted by ``${\rm \Omega}$". 
The background contribution due to the LCC effect has a similar characteristic structure 
as the above~\cite{Pratt2010,PrattSorren:2011}.
$\kappa_{112}^{\rm TMC}$ has been evaluated for $0-50\%$ Au+Au collisions at 200 GeV,
with the $v_2$ measurements by the PHOBOS collaboration~\cite{PHOBOS1,PHOBOS2}.
Figure~5(a) shows a good agreement between the $\kappa_{112}^{\rm TMC}$ thus obtained and the 
$\kappa_{112}$ from the background-only AMPT calculations.
On the other hand, STAR $\kappa_{112}$ data are significantly higher than the estimated background, except in the most central collisions, where the CME signal is expected to be minimal owing to the vanishing magnetic field. 

The $\gamma_{123}$ can be rewritten in a similar way as Eq.(17),
\begin{eqnarray}
\gamma_{123} &=& \langle \cos( \phi_{\alpha} + 2\phi_{\beta} - 3\Psi_{3} ) \rangle / Res\{\Psi_{3}\}\nonumber \\
&=& \delta\cdot v_3 + \langle \langle \cos(\phi_{\beta} - \phi_{\alpha})\cos(2\phi_{\beta} - 2\Psi_{3} )  \rangle \rangle/Res\{\Psi_{3}\} \nonumber \\
& & +\langle \langle \sin(\phi_{\beta} - \phi_{\alpha})\sin(2\phi_{\beta} - 2\Psi_{3} )  \rangle\rangle/Res\{\Psi_{3}\},
\end{eqnarray}
with the two cumulant terms also adding up. However, there is no obvious reason to expect $\kappa_{123} \approx \kappa_{112}$.
A constant fit over $0-80\%$
centrality range gives $\kappa_{123} = 2.17 \pm 0.15$,
which is higher than $\kappa_{112}$ with a $5.7\sigma$ significance with the current AMPT statistics. 
In other words, the absence of the CME does not validate the equality between $\kappa_{123}$ and $\kappa_{112}$. Note that the $\kappa_{123}$ values from AMPT can describe the experimental data reasonably well, and thus support the idea that $\gamma_{123}$ only contains background contributions.
Even so, we should not regard $\kappa_{123}$ as a background measure for $\kappa_{112}$.
The background scenario generated by AMPT can well describe the features of $v_2$, $v_3$ and $\Delta\delta$ in STAR data, but expects $\kappa_{112}$ and $\kappa_{123}$ to be different.

\section{Summary}

We extend the previous CME-related observables ($\gamma_{112}$ and $\gamma_{123}$) to $\gamma_{132}$ to quantitatively investigate the background contributions in these azimuthal correlators. With certain assumptions in the pure-background 
scenario, $\gamma_{132}$ and $\gamma_{123}$, when propoerly normalized, could serve as data-driven background estimates
for $\gamma_{112}$, the latter containing both signal and background contributions.
However, the usefulness of $\gamma_{132}$ and $\gamma_{123}$,
or $\kappa_{132}$ and $\kappa_{123}$, has to be tested with simulations from a realistic event generator.
We have employed a charge-conserving version of AMPT to perform this test for Au+Au collisions at 200 GeV.

In the absence of the CME, the AMPT calculations reasonably well reproduce the experimental data of the centrality dependence of $v_2$, $v_3$ and $\Delta\delta$, making a promising candidate for the background estimation for the CME measurements. The simulated $\Delta\gamma_{112}$, $\Delta\gamma_{132}$ and $\Delta\gamma_{123}$ all reveal positively finite values in the centrality range under study, and the centrality dependence resembles the CME expectation. This warrants a careful investigation of the background contributions. On the other hand, the STAR $\Delta\gamma_{112}$ results are significantly higher than AMPT, and possible CME contributions can not be ruled out by our background studies. 

After normalization with $v_2$ and $\Delta\delta$, 
$\kappa_{132}$ from AMPT seems to be constant over the $0-80\%$ centrality range, and consistent with unity. This 
observation indicates that some pertinent correlations in $\gamma_{132}$ are symmetric between in-plane and out-of-plane, and cancel each other.
$\kappa_{132}$ serves as a first-order background estimate for $\kappa_{112}$, and $v_2\cdot\delta^{\rm OS(SS)}$ provides a prediction for $\gamma_{132}^{\rm OS(SS)}$ in data. The background contribution in $\kappa_{112}$ has been estimated with both AMPT and $v_2$ measurements from PHOBOS, the latter of which takes into account collective motion and TMC.
Both results show very weak centrality dependence, and are consistently higher than $\kappa_{132}$ by about $30\%$.
Conversely, the STAR $\kappa_{112}$ data demonstrate a strong centrality dependence, and are significantly higher than the estimated background, except in the most central collisions. These features of STAR data deserve further investigations in search of the CME in heavy-ion collisions. 

With the current AMPT statistics, $\kappa_{123}$ is higher than $\kappa_{112}$ with a $5.7\sigma$ significance. Therefore, in the pure-background scenario, we should not expect an equality between $\kappa_{123}$ and $\kappa_{112}$.
There is no indication in our simulation that $\kappa_{123}$ is a reliable data-driven background measure for $\kappa_{112}$, though experimentally their magnitudes are close to each other. The latter fact remains a challenge to both the CME and the background scenarios. This calls for future works beyond the pure-background scenario of AMPT.
The recently developed Anomalous-Viscous Fluid Dynamics (AVFD)~\cite{summary_39,summary_40} implements the CME signal together with realistic background contributions, and makes a promising tool for this purpose.

{\bf Acknowledgments:} The authors thank Zi-Wei Lin and Guo-Liang Ma for providing the AMPT code. We are grateful to Aihong Tang and Jinfeng Liao for helpful
communications and discussions.
The research is supported by the National Natural Science Foundation of China under the Grant 11835002 and by the U.S. Department of Energy, Office of Nuclear Physics, under the Grant DE-FG02-88ER40424.

\section{Appendix A}

We want to elaborate a background scenario with flowing resonances,
e.g., $\rho$ mesons that decay into $\pi^+$ and $\pi^-$. In the absence of the CME, the $\gamma_{112}$ correlator for all opposite-sign pion pairs has the following background contribution
\begin{eqnarray}
\gamma_{112}^{\rm BG} &=& \langle \cos( \phi_{\alpha} + \phi_{\beta} - 2\Psi_{\rm RP} ) \rangle \nonumber \\
&=& \langle \cos( \phi_{\alpha} + \phi_{\beta} - 2\phi_{\rho} + 2\phi_{\rho} - 2\Psi_{\rm RP} ) \rangle \nonumber \\
&\approx& \frac{f_{\rho/\pi^{\pm}}}{N_{\pi^{\pm}}}
\langle \cos(\phi_{\alpha} + \phi_{\beta} - 2\phi_{\rho}) \rangle v_{2,\rho},
\end{eqnarray}
where $f_{\rho/\pi^{\pm}}$ is the fraction of $\rho$-decayed pions.
In the same way, the $\gamma_{132}$ correlator contains a similar background
\begin{equation}
\gamma_{132}^{\rm BG} = \frac{f_{\rho/\pi^{\pm}}}{N_{\pi^{\pm}}}
\langle \cos(\phi_{\alpha} - 3\phi_{\beta} + 2\phi_{\rho}) \rangle v_{2,\rho}.    
\end{equation}
Since both decay pions are boosted by the parent $\rho$ meson, we may consider a crude assumption that $\phi_{\alpha} \approx \phi_{\beta}$, which leads to
$\langle \cos(\phi_{\alpha} + \phi_{\beta} - 2\phi_{\rho}) \rangle \approx \langle \cos(\phi_{\alpha} - 3\phi_{\beta} + 2\phi_{\rho}) \rangle$, or simply
$\gamma_{112}^{\rm BG}\approx\gamma_{132}^{\rm BG}$.
This simplified picture helps us understand how $\kappa_{132}$ can serve as a first-order background estimate for $\kappa_{112}$ in a special physics process.

A similar derivation for $\gamma_{123}$ in this scenario yields
\begin{equation}
\gamma_{123}^{\rm BG} = \frac{f_{\rho/\pi^{\pm}}}{N_{\pi^{\pm}}}
\langle \cos(\phi_{\alpha} + 2\phi_{\beta} - 3\phi_{\rho}) \rangle v_{3,\rho}.    
\end{equation}
Once we require $\phi_{\alpha} \approx \phi_{\beta}$, the parent $\rho$ meson should also go along the same direction, making
$\langle \cos(\phi_{\alpha} + \phi_{\beta} - 2\phi_{\rho}) \rangle \approx \langle \cos(\phi_{\alpha} + 2\phi_{\beta} - 3\phi_{\rho}) \rangle \approx 1$,
or simply $\gamma_{112}^{\rm BG}/v_{2,\rho}\approx\gamma_{123}^{\rm BG}/v_{3,\rho}$. Therefore $\kappa_{123}$ may also provide a background estimate for $\kappa_{112}$ in this special case. 

We may refine the picture by  considering $\phi_\alpha = \phi_\rho + \Delta\phi$ and $\phi_\beta = \phi_\rho - \Delta\phi$, where $\Delta\phi$ is a small angle.  Then the core terms in $\gamma_{112}^{\rm BG}$, $\gamma_{132}^{\rm BG}$ and $\gamma_{123}^{\rm BG}$ become $\langle\cos(0)\rangle=1$, $\langle\cos(4\Delta\phi)\rangle$ and $\langle\cos(\Delta\phi)\rangle$, respectively. 
In reality, the aforementioned assumption is oversimplified, and $\Delta\phi$ is not always small. Therefore, we need a more realistic model such as AMPT to perform the background estimation.

\end{document}